\documentclass[prb,superscriptaddress,showpacs,twocolumn]{revtex4}

\usepackage{amsmath,amsfonts,mathtools,bm,braket}
\usepackage{cancel,MnSymbol}

\usepackage{graphicx}
\graphicspath{{images/}}

\usepackage{hyperref}
\hypersetup{
	colorlinks   = true, 
	urlcolor     = blue, 
	linkcolor    = blue, 
	citecolor   = blue 
}

\newcommand{\abs}[1]{\left|#1\right|}

\renewcommand{\vec}[1]{{\boldsymbol #1}}

\renewcommand{\d}{\mathrm{d}}

\renewcommand{\S}{{\cal S}}

\begin{document}
	
\title{Electrons at the monkey saddle: a multicritical Lifshitz point}

\author{A. Shtyk}
\affiliation{Department of Physics, Harvard University, Cambridge, MA 02138, USA}
\author{G. Goldstein}
\affiliation{Cavendish Laboratory, University of Cambridge, Cambridge, CB3 0HE, United Kingdom}
\author{C. Chamon}
\affiliation{Department of Physics, Boston University, Boston, MA, 02215, USA}

\begin{abstract}
We consider 2D interacting electrons at a monkey saddle with
dispersion $\propto p_x^3-3p_xp_y^2$. Such a dispersion naturally
arises at the multicritical Lifshitz point when three van Hove saddles
merge in an elliptical umbilic elementary catastrophe, which we show
can be realized in biased bilayer graphene. A multicritical Lifshitz
point of this kind can be identified by its signature Landau level
behavior $E_m\propto (Bm)^{3/2}$ and related oscillations in
thermodynamic and transport properties, such as de Haas-van Alphen and
Shubnikov-de Haas oscillations, whose period triples as the system
crosses the singularity. We show, in the case of a single monkey
saddle, that the non-interacting electron fixed point is unstable to
interactions under the renormalization group flow, developing either a
superconducting instability or non-Fermi liquid features. Biased
bilayer graphene, where there are two non-nested monkey saddles at the
$K$ and $K^\prime$ points, exhibits an interplay of competing
many-body instabilities, namely $s$-wave superconductivity, ferromagnetism, and spin-
and charge-density wave.
\end{abstract}

\maketitle

\section{Introduction}

Systems of two-dimensional (2D) electrons close to van Hove (vH)
singularities\cite{furukawa,dzyaloshinskii,kane92,ziletti,gonzalez,gopalan,menashe,katsnelson,kapustin}
are of interest because of their displayed logarithmic enhancement of
the electron density of states (DoS), which translates into a
propensity to many-body instabilities\cite{furukawa}. Among many
exciting possibilities opened by proximity to vH singularities is that
unconventional $d+id$ chiral superconductivity could occur in strongly
doped graphene monolayer\cite{nandkishore}.

\begin{figure}[t!]
\center{\includegraphics[width=1\linewidth]{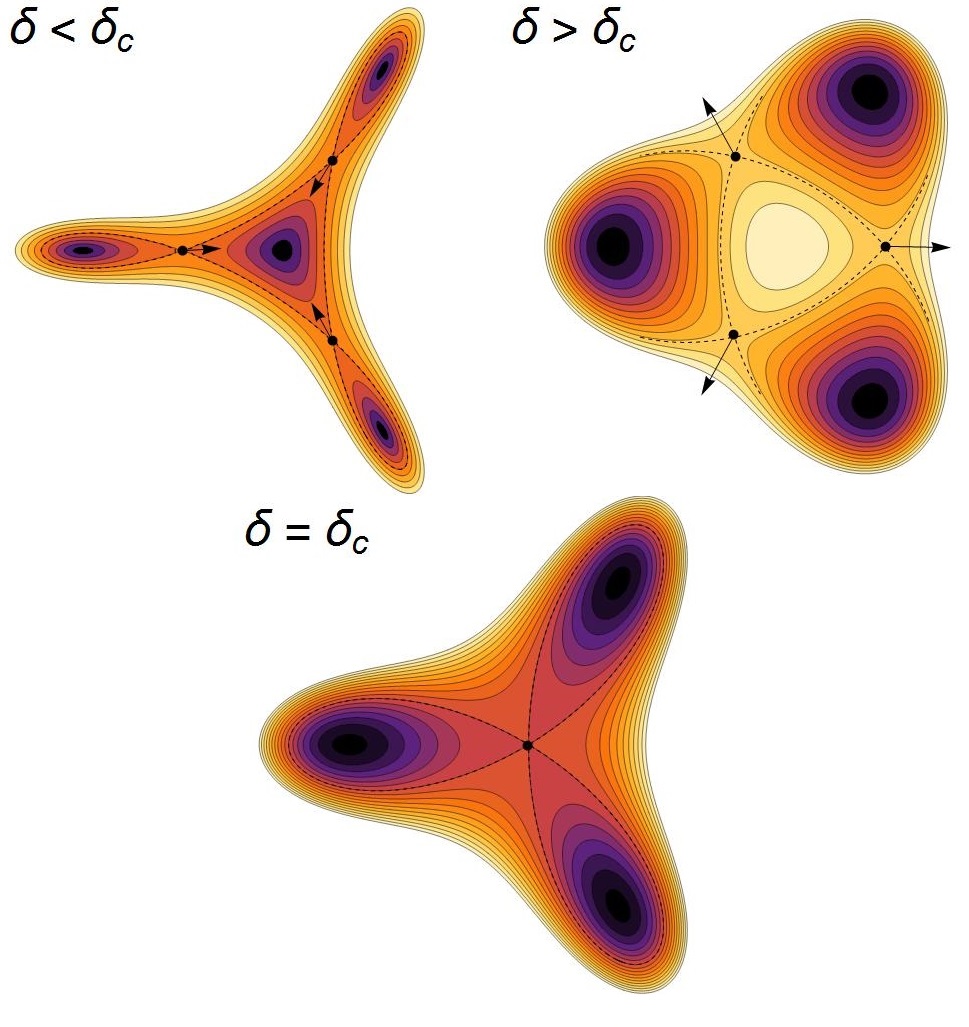}} 
\caption{Pictorial representation of Fermi surface families in a
  biased bilayer graphene system for three different values of the
  interlayer voltage bias $\delta$. Three van Hove saddles with
  dispersions $\propto (p_x^2-p_y^2)$ are shown with black dots
  ($\delta\neq\delta_c$) while arrows indicate their displacement upon
  increasing the value of $\delta$. At the critical value of the bias
  $\delta_c$ they merge into a monkey saddle
  $\propto(p_x^3-3p_xp_y^2)$ that closes into a trifolium-shaped Fermi
  surface.	
	}
	\label{fig:FS_families}
\end{figure}

The transition of the Fermi level through a vH singularity can be
interpreted essentially as a Lifshitz transition of a neck-narrowing
type\cite{ghamari}, wherein two disconnected regions of the Fermi
surface (FS) merge together. Alternatively, if the touching occurs at
the edge of the Brillouin zone, as it happens for the square lattice,
it may be interpreted as a FS turning inside out (from electron-like
to hole-like). A multicritical Lifshitz point (MLP) arises as both a
crossing of several Lifshitz transition lines, and as a singularity in
the electronic dispersion $\xi(\vec{p})$. MLPs of bosonic type have
been analyzed and classified in the context of phase transitions,
where terms in the free-energy-density functional with higher-order
derivatives of an order parameter, say the magnetization, need to be
kept at special points in the phase
diagram~\cite{hornreich,aharony,thede}. Yet, MLPs of fermionic type,
with a singularity in the fermionic dispersion $\xi(\vec{p})$, have
been largely unexplored, only in a scenario involving Majorana
fermions and spin liquids~\cite{biswas} where the monkey saddle was
produced because of symmetries of the low energy Hamiltonian as
opposed to a merging of several vH singularities.

\begin{figure}[t]
	\center{\includegraphics[width=1\linewidth]{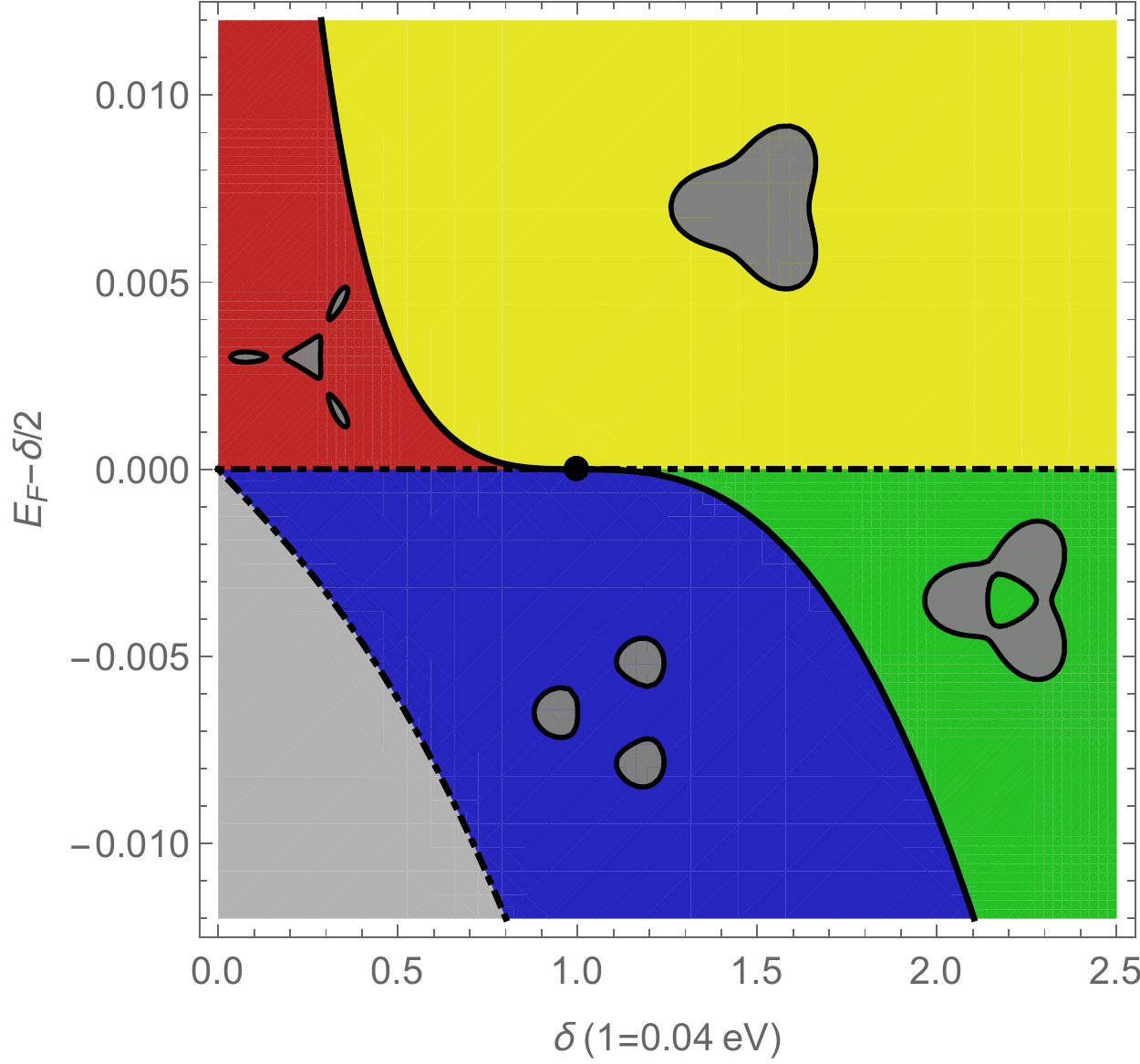}} 
	\caption{There are four phases with different Fermi surface topologies
		in biased bilayer graphene. They are separated by two lines of phase
		transitions, one of a band-edge type (dash-dotted) and the other of
		the van Hove or equivalently a neck-narrowing type (solid). The
		multicritical Lifshitz point is located at the crossing of these two
		lines. In the gray area the Fermi level lies within the gap with no
		FS. Note different scales for the voltage $\delta$ and the Fermi
		energy $E_F$.}
	\label{fig:phases}
\end{figure}

In this paper we study fermionic MLPs, using biased bilayer graphene
(BLG) as a concrete example of a physical realization. In the case of
BLG, three vH saddles merge into a monkey saddle at critical value of
the interlayer voltage bias (see
Figs.~\ref{fig:FS_families},\ref{fig:phases}). Mathematically, the
monkey saddle is a genuine mathematical singularity with a degenerate
quadratic form as opposed to vH saddle, which is not a true
singularity in a mathematical sense, having a non-degenerate quadratic
form of the $(+-)$ signature, $\propto p_x^2-p_y^2$. Physically, we
identify key differences between the case of a MLP and that of the
usual vH singularity. First, the monkey-saddle-like dispersion
$\propto p_x^3-3p_xp_y^2$ at the MLP exhibits a stronger, power-law
divergence in the DoS and thus leads to even stronger many-body
instabilities, with higher transition temperatures as a result. These
stronger DoS divergences greatly simplify the renormalization group
(RG) analysis of the problem, yielding a super-renormalizable
theory. We find that the non-interacting electron fixed point is
unstable to interactions, developing either a superconducting
instability or non-Fermi liquid behavior. In the case of BLG, which
has two non-nested monkey saddles at the $K$ and $K'$ points,
interactions lead to instabilities to $s$-wave superconducting state, ferromagnetism,
spin-, and charge density wave, depending on the nature of interactions. Second,
the monkey saddle possesses a signature Landau level (LL) structure
with energy levels $E_m\propto(Bm)^{3/2}$. In addition, oscillations
in different thermodynamic and transport properties, such as de
Haas-van Alphen and Shubnikov de Haas oscillations, for example, are
sensitive to the presence of the multicritical point. The monkey
saddle can be identified by the scaling of the period of these
oscillations with the Fermi energy as $\Delta(1/B)\propto E_F^{2/3}$
and with an abrupt tripling of the period as Fermi level goes from
below to above the saddle, due to a change of the FS topology.

The presentation of the results in the paper is organized as
follows. In Sec.~\ref{sec:dispersion} we present how the monkey saddle
arises in voltage-biased BLG. We show how four different FS topologies
can be attained by varying the bias voltage and the chemical
potential, and identify the MLP in the phase diagram as the location
where these four different phases meet at a point. There we also
discuss the nature of the divergence in the density of states for the
monkey saddle dispersion. In Sec.~\ref{sec:LL} we obtain the energies
of the quantized Landau orbits within a quasiclassical approximation,
and present arguments for the period tripling of the magnetic
oscillations as the system undergoes a FS topology change; these
features may serve as clear experimental telltales of the MLP in
BLG. In Sec.~\ref{sec:RG1} we present an RG analysis of the case when
interactions are present in a system with an isolated monkey saddle,
where we show that the system is either unstable to superconductivity
or flows to a non-Fermi liquid, depending on the sign of the
interactions. The RG analysis for the case of BLG with two monkey
saddles at the $K$ and $K'$ points is studied in Sec.~\ref{sec:RG2},
where we discuss the possible instabilities of the system. We close
the paper by summarizing the results and discussing open problem in
Sec.~\ref{conclusions}.


\section{Hamiltonian and dispersion}
\label{sec:dispersion}

Here we explicitly show how the monkey saddle arises in BLG. We
consider AB-type stacked BLG, with the layers labeled by $1$ and $2$,
and the two sublattices within each layer labeled by $A$ and $B$. The
spinor representing the electronic amplitudes is chosen in the order
$({A1}, {B1}, {A2}, {B2})$. We consider an extended tight-binding
model that includes next-nearest neighbor hopping, where the
Hamiltonian of the system linearized near the $K$ point
is~\cite{koshino}
\begin{align}
\label{H_0}
	\check{H}_0=
	\begin{pmatrix}
		\frac{1}{2}V & vp_- & 0 & v_3p_+
		\\
		vp_+ & \frac{1}{2}V & \gamma_1 & 0
		\\
		0 & \gamma_1 & -\frac{1}{2}V & vp_-
		\\
		v_3p_- & 0 & vp_+ & -\frac{1}{2}V
	\end{pmatrix}.
\end{align}
Here $v$ is the band velocity of monolayer graphene,
$\gamma_1=0.4\text{ eV}$ is an interlayer coupling constant and
\mbox{$v_3\approx0.1v$} describes trigonal warping that arises as a
result of the next-nearest-neighbor hopping. $V$ is an interlayer
voltage bias and $p_\pm=p_x\pm ip_y$ is the momentum. BLG has four
energy bands and in this paper we are focused solely on the lowest
upper band with an electron dispersion\cite{varlet}
\begin{equation}
	\begin{split}
	\xi^2(\vec{p})=
	\frac{V^2}{4}\left(1-2\frac{v^2p^2}{\gamma_1^2}\right)^2
	+
	v_3^2p^2
	+\dots
	\\
	\dots+
	2\frac{v_3v^2}{\gamma_1}p^3\cos3\phi
	+
	\frac{v^4p^4}{\gamma_1^2}.
	\end{split}
\end{equation}
For voltage biases $V$ of the order of the trigonal warping energy scale $\gamma_1$ the $\propto p^4$ contribution arising from the first term can be safely neglected.
It is convenient to introduce dimensionless variables, redefining energies as $\xi\rightarrow (v_3\gamma_1/v)\xi$ and momenta as $\vec{p}\rightarrow (v_3\gamma_1/v^2)\vec{p}$, 
\begin{align}
\label{xi}
	\xi^2(\vec{p})=(\delta/2)^2+u_3^2\left[(1-\delta^2)p^2+2p^3\cos3\phi+p^4\right],
\end{align}
where $u_3\equiv v_3/v\approx0.1$ is a dimensionless measure of the
warping strength and $\delta\equiv V/(v_3\gamma_1/v)$. The dispersion
near the $K^\prime$ point can be obtained from the one near the $K$
point by inversion, $\vec{p}\rightarrow-\vec{p}$.

Unlike in the case of a monolayer graphene, where the warping merely
distorts the Dirac cone with low-energy dispersion unaffected, BLG
behaves in a very different way. In the absence of interlayer voltage
bias, the trigonal warping destroys the parabolic dispersion, breaking
it down into four Dirac cones. A non-zero interlayer voltage $V$ gaps
out these Dirac cones while also gradually inverting the central
electron pocket into a hole-like pocket at the critical value of the
bias $V_c=(v_3/v)\gamma_1$ ($\delta_c=1$ in dimensionless units
introduced above). This critical value of the bias marks a singularity
in the electronic dispersion $\xi(\vec{p})$.

At the subcritical interlayer voltage bias $\delta<1$ the electronic dispersion $\xi(\vec{p})$ has seven extremal points, four electronic pockets and three vH saddle points. While the three outer electronic pockets are robust and are present at all voltage biases, the central extremum and three vH saddle points merge at the critical voltage falling apart again into three saddles and a hole-like pocket at the supercritical bias $\delta>1$, see Fig. \ref{fig:FS_families}.

In the vicinity of the singular point the electronic dispersion behavior is governed by the lowest powers of the momentum:
\begin{align}
	\xi(\vec{p})\propto\underbrace{(1-\delta^2)p^2}_{\text{Pert(2,1)}}+\underbrace{p^3\cos 3\phi}_{\text{CG(2)}}.
\end{align}
This momentum behavior corresponds exactly to the symmetry-restricted
elliptic umbilic elementary catastrophe ($D_4^-$ within $ADE$
classification)~\cite{Arnold}. From the point of view of the
catastrophe theory the cubic term $p^3\cos3\phi\equiv \text{CG(2)}$ is
a catastrophe germ defining the nature of the singularity in
$\xi(\vec{p})$ function, while the quadratic term
$(1-\delta^2)p^2\equiv \text{Pert(2,1)}$ is a lattice-symmetry
restricted perturbation, with one parameter $\delta$, which regularizes
the singularity. Qualitatively the behavior of the system can be
viewed as a bifurcation of a monkey saddle $p^3\cos3\phi\equiv
p_x^3-3p_xp_y^2$ into three vH(ordinary) saddles and a
maximum/minimum:
\begin{equation}
	\underbrace{p_x^3-3p_xp_y^2}_{\text{monkey saddle}} \longleftrightarrow 3\times \underbrace{(p_x^2-p_y^2)}_{\mathclap{\text{vH saddle}}}+1\times \underbrace{p^2}_{\mathclap{\text{e/h pocket}}}.
\end{equation}

\subsection{Strong density of states divergence}
\label{sec:DoS}

The monkey saddle leads to a strong IR divergence in the DoS. While
the vH saddle has a logarithmic DoS, any generic higher order saddle
$\xi(\vec{p},n)=ap^n\cos n\phi$ has a power-law divergence in the
DoS. To obtain the DoS for a higher order saddle, it is convenient to
work on generalized hyperbolic coordinates $(\xi,\eta)=a(p^n\cos
n\phi, p^n\sin n\phi)$ (where $n=1,2$ correspond to polar and
hyperbolic coordinates, respectively). The dispersion of the saddle is
given by the $\xi$ variable, while $\eta$ plays the role of the
hyperbolic angle, parametrizing displacements along the FS. The
density of states is given by
\begin{equation}
\begin{split}
	\nu(\xi,n)
	=
	\oint_{FS}\frac{(\mathrm{d}\vec{p})}{\mathrm{d}\xi}=\frac{a^{-2/n}}{(2\pi)^2n}\int_{-\infty}^{+\infty}\frac{\d\eta}{(\xi^2+\eta^2)^{\frac{n-1}{n}}}
	\\
	=
	\frac{a^{-2/n}}{4n\pi^{3/2}}\frac{\Gamma\left(\frac{1}{2}-\frac{1}{n}\right)}{\Gamma\left(1-\frac{1}{n}\right)}\xi^{-\frac{n-2}{n}},
\end{split}
\label{DoS}
\end{equation}
where the $(\d\vec{p})\equiv d^2p/(2\pi)^2$, and we set Planck's constant
to unit ($\hbar=1$).

\begin{figure}[t]
	\center{\includegraphics[width=.7\linewidth]{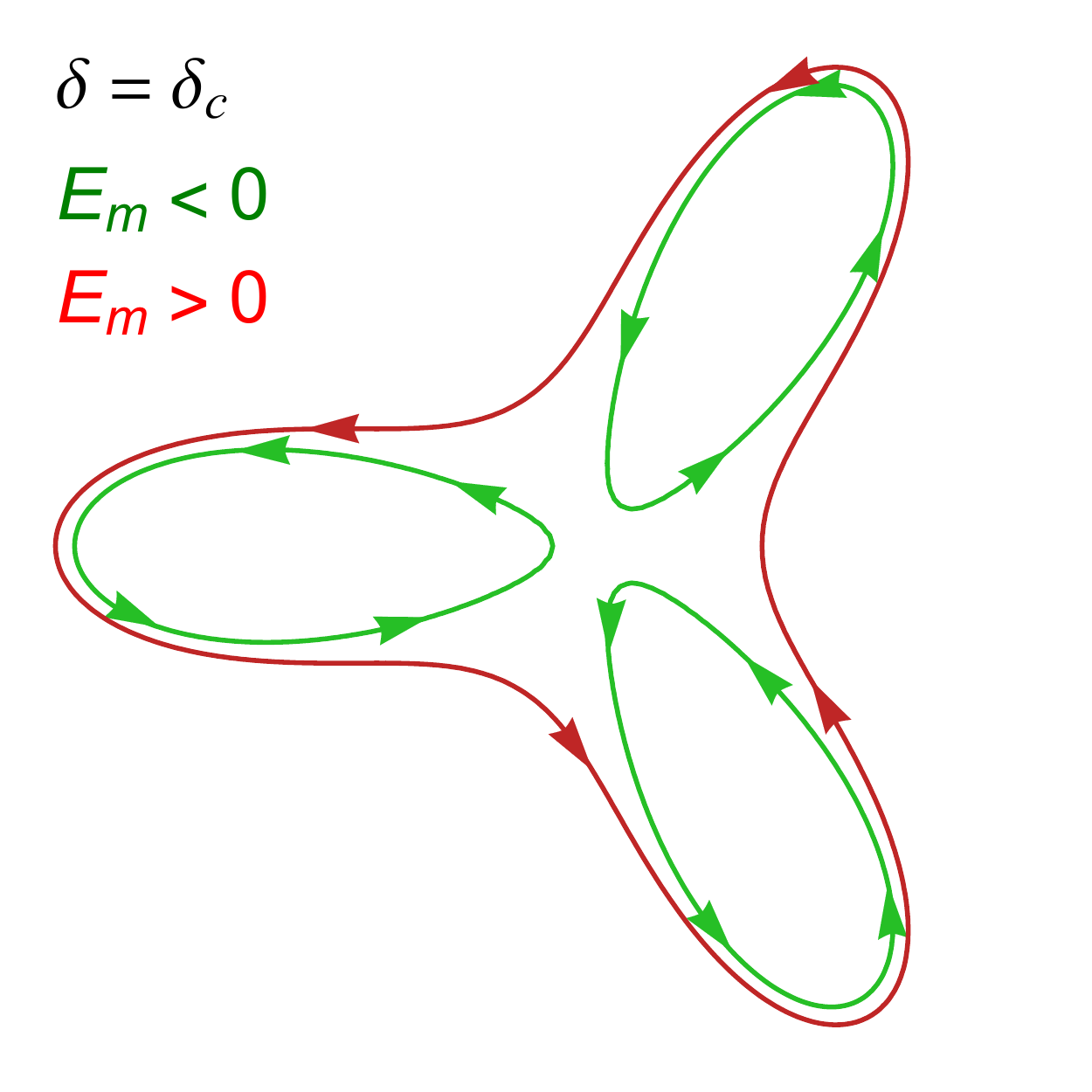}} 
	\caption{Quasiclassical LL orbits in momentum space for energies
		slightly below and slightly above the monkey saddle (and critical
		voltage bias). The number of connected FS components changes from
		three to one as the Fermi level crosses zero. }
	\label{fig:LL_orbits}
\end{figure}

\subsection{Fermi surface topology phase diagram}
\label{sec:FS}

The electron FS at a given Fermi energy is defined as a cross-section
of the electron dispersion $\xi(\vec{p},\delta)=E_F$.  There are four
distinct Fermi surface topology phases within the $(\delta,E_F)$ plane
(see Fig.~\ref{fig:phases}). All of them have the same three-fold
symmetry but can be discerned by their topological invariants, the
number of connected components and the number of holes. Namely, in our
case the four phases can be labeled uniquely by the first two Betti
numbers of their FS $(b_{0},b_{1})$ as (1,0), (4,0), (3,0), and (1,1).

These four phases are separated by two lines of topological phase
transitions. One of the lines is of a weaker, band-edge transition
type, while another is of a stronger vH type (the former has a jump in
the DoS while the latter has a log-divergence). The multicritical
Lifshitz point lies at the intersection of these two lines.

\section{Magnetic oscillations at the monkey saddle}
\label{sec:LL}

Within a quasiclassical approximation, the LLs can be obtained by
quantization of the area enclosed by quasiparticle orbit in momentum
space,
\begin{equation}
	\int(d\vec{p})=\frac{m}{2\pi l_B^2},
\end{equation}
where $l_B=\sqrt{c/eB}$ is a magnetic length and $m$ is the LL
index. For a system tuned exactly to the monkey saddle (or any higher
order saddle), the behavior is dominated by the singularity itself, so
that
\begin{equation}
\label{LL}
\begin{split}
  \int_0^{E_m}\nu(\xi)d\xi=\frac{1}{8\pi^{1/2}}\frac{\Gamma\left(\frac{1}{2}-\frac{1}{n}\right)}{\Gamma\left(1-\frac{1}{n}\right)}\left(\frac{E_m}{a}\right)^{\frac{2}{n}}
\\
\implies
E_m=
\alpha\left(\frac{a}{l_B^n} \right)m^{n/2}\propto(Bm)^{n/2}
\end{split}
\end{equation}
with a numerical coefficient
\begin{equation}
	\alpha=
	\left(4\sqrt{\pi}\frac{\Gamma\left(1-\frac{1}{n}\right)}{\Gamma\left(\frac{1}{2}-\frac{1}{n}\right)}\right)^{\frac{n}{2}}\underset{(n=3)}{=}2.27.
\end{equation}

As always, LLs imply oscillations of various transport and
thermodynamic properties in an applied magnetic field. Since such
oscillations happen as LLs cross the Fermi level of the system. At the
critical voltage bias $\delta_c=1$ but with a small positive detuning
from the energy of the saddle point, {\it i.e.}, $E_F$ slightly higher
than $\delta_c/2$, we can see from Eq.(\ref{LL}) that we have a
periodicity in inverse magnetic field with a period
\begin{equation}
	\label{B_period}
	\Delta\left(\frac{1}{B}\right)=\frac{e\hbar}{c}\left(\frac{E_F}{\alpha a}\right)^{2/n},
\end{equation}
where we reinserted Planck's constant $\hbar$.

Eqs.(\ref{LL},\ref{B_period}) are given for positive LL energies, when
$E_F$ slightly higher than $\delta_c/2$, and the FS consists of one
connected component, see
Figs.~\ref{fig:FS_families},\ref{fig:phases}. The situation is
different for negative energies, when when $E_F$ slightly lower than
$\delta_c/2$ and the Fermi surface has three disconnected
components. In this case the LLs are triply degenerate (on top of the
valley degeneracy), and are three times as sparse,
\begin{equation}
	E_{-m}=-\alpha al_B^{-n}(3m)^{n/2},
\end{equation}
and oscillations period in inverse magnetic field is three times
smaller as well. (All equations above are for spinless electrons: in a
real system Zeeman splitting should be taken into account as well.)

The tripling of the periodicity of oscillation is a telltale of the
Fermi surface topology change, and can be viewed physically as
follows. The area of the Fermi surface is not very different slightly
before or slightly after it undergoes the topology change. At the
critical point, the area that fits just one electron orbit is brought
inside the Fermi surface upon insertion of a flux quantum. When there
is a single surface, one can indeed fit a physical electron within
that orbit. However, when the Fermi surface contains the three
pockets, the additional area brought inside each pocket due to a
single flux quantum insertion is only 1/3 of what is needed to fit one
electron. If there were quasiparticles with charge 1/3, then they
could fill separately the area in the three pockets; but there are no
such particles in the system. Hence, the flux periodicity is tripled
when the Fermi surfaces are disconnected, as one can only add a full
electron at each pocket, requiring the addition of three flux
quanta. This is the physical origin of the period tripling.

\begin{figure}[t]
	\begin{minipage}[h]{0.49\linewidth}
		\center{\includegraphics[width=1\linewidth]{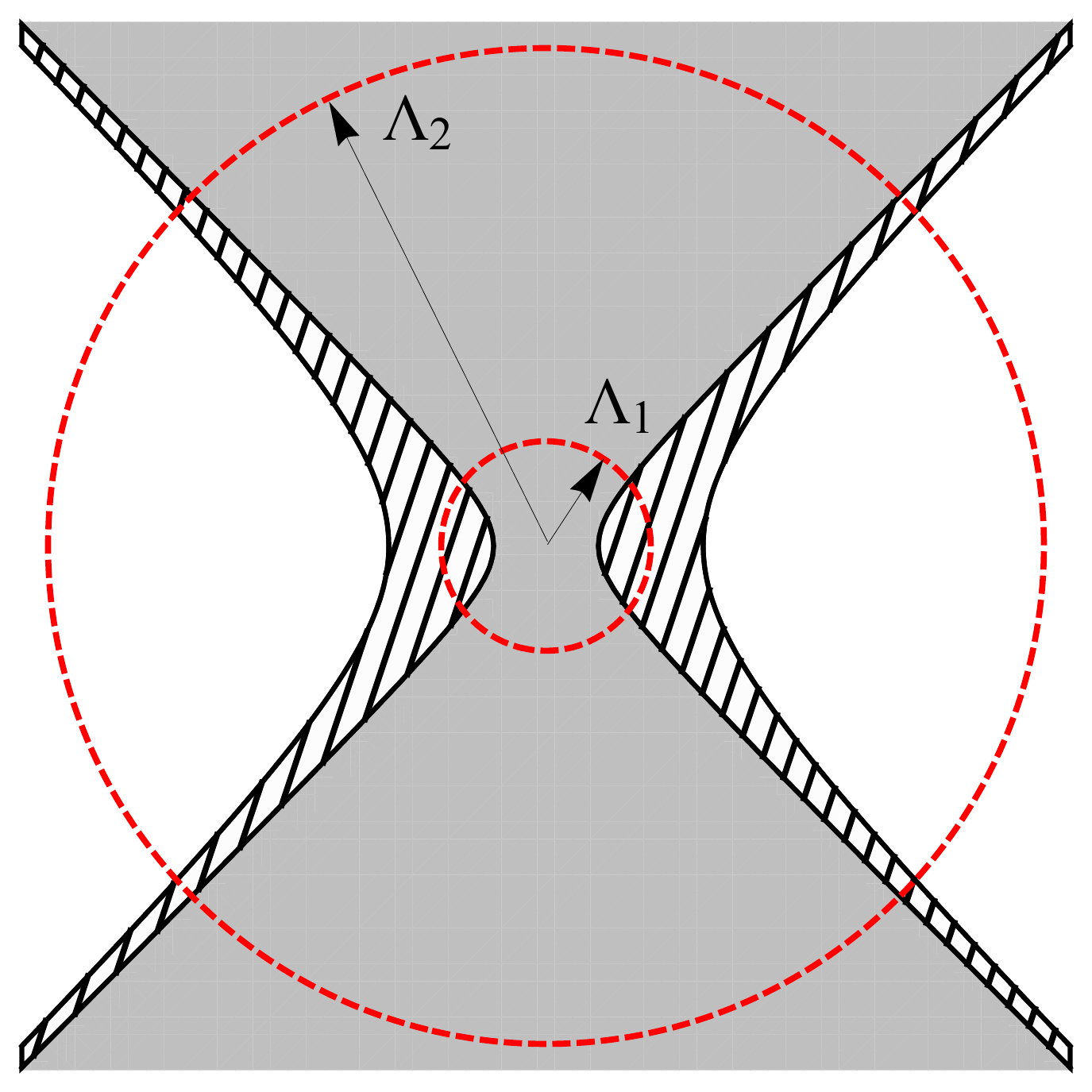}}
	\end{minipage}
	\hfill
	\begin{minipage}[h]{0.49\linewidth}
		\center{\includegraphics[width=1\linewidth]{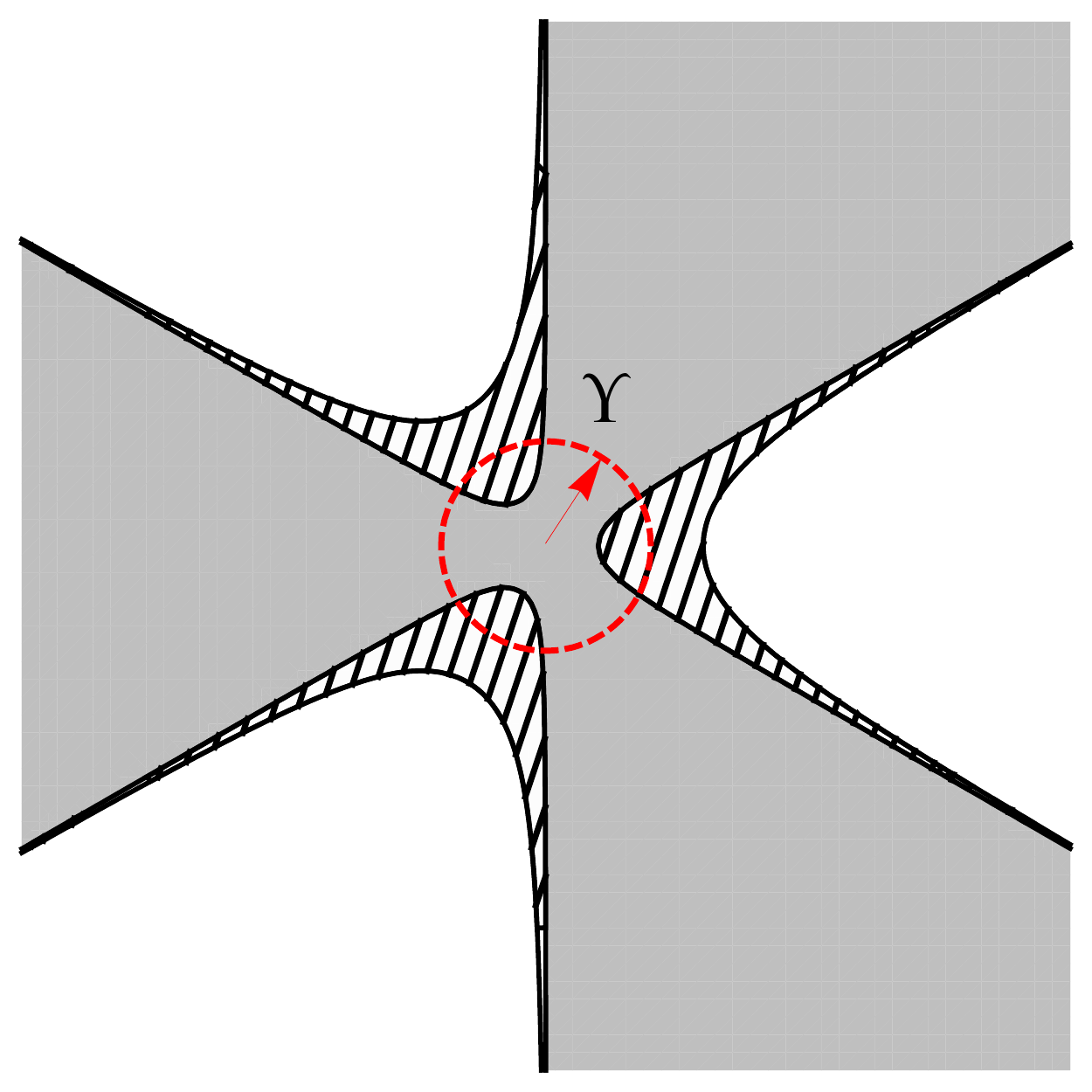}}
	\end{minipage}
	\caption{\textit{Left}: A Fermi surface near a van Hove saddle calls
		for a two-cutoff RG scheme. The grey area represents occupied
		electron states. The hatched region of the phase space corresponds
		to a step $d\xi$ in electron energy. Normally, one cutoff
		$d\Lambda_1\sim d\xi$ is sufficient, but here we see that the
		logarithmic DoS at the van Hove saddle together with an open
		hyperbolic Fermi surface lead to tails of the hatched region that
		reach out to the rest of the Fermi surface away from the van Hove
		saddle. The purpose of the second cutoff $\Lambda_2$ is to cut these
		tails and isolate van Hove saddle.
		\newline	
		\textit{Right}: No second cutoff is needed at the
		monkey saddle.}
	\label{fig:RG_cutoff}
\end{figure}

\section{RG flow at the monkey saddle}
\label{sec:RG1}

Here we analyze a single monkey saddle within a one-loop RG
framework. Assuming short-range interaction, an electron action is
given by
\begin{align}
\S=\int(\d \tau\d\vec{r})\left[\psi^\dagger[\partial_\tau-\xi(-i\vec{\nabla})+\mu]\psi-\frac{g}{2}(\psi^\dagger\psi)^2\right]
\end{align}
with interaction
\begin{equation}
\frac{g}{2}(\psi^\dagger\psi)^2=g(\psi^\dagger_{\uparrow}\psi^\dagger_{\downarrow}\psi_{\downarrow}\psi_{\uparrow}).
\end{equation}

We focus on the system tuned exactly to the monkey saddle, so that the
dispersion is determined by the catastrophe germ
$\xi(\vec{p})=p^3\cos3\phi$ and the non-singular part of FS is
irrelevant (see Fig.~\ref{fig:RG_cutoff}). Tree-level RG involves
rescaling of frequency and momenta as
\begin{equation}
\omega\rightarrow s^{-1}\omega,\,\vec{p}\rightarrow s^{-1/3}\vec{p},\,\psi\rightarrow s^{-1/3}\psi,
\end{equation}
and results in the interaction constant scaling as 
\begin{equation}
\label{g_scaling}
g\rightarrow gs^{+1/3},
\end{equation}
entailing super-renormalizability of the theory.

Super-renormalizability brings crucial simplifications with respect to
the case of the ordinary vH saddle: while the separation of the saddle
from the non-singular part of the FS requires two cut-offs in the case
of vH singularities ($n=2$), it requires only one cut-off for higher
order singularities ($n>2$), see Fig.~\ref{fig:RG_cutoff}. This
difference can be traced back to the behavior of DoS obtained in
Eq.(\ref{DoS}). In the case of the vH saddle ($n=2$), the integral
over the angle-like variable $\eta$ diverges logarithmically,
requiring an additional cut-off in the problem that is interpreted as
a Fermi velocity cut-off in Refs.~\onlinecite{ghamari,kapustin}. In
contrast, for any higher-order saddle with $n>2$, the DoS at a given
energy is well-defined and is determined solely by the saddle and does
not require a large momentum cut-off. This means that for $n>2$ the
theory is free of UV divergences and contains only (meaningful) IR
divergences that are regularized by temperature $T$ and chemical
potential $\mu$.

We introduce a dimensionless coupling constant in a natural way as
\begin{equation}
\label{lambda}
\lambda(\Upsilon)=\nu(\Upsilon) g(\Upsilon),	
\end{equation}
with a smooth \textit{infrared} cutoff $\Upsilon$ that we take to be
either $\mu$ or $T$, so that the beta function for the dimensionless
coupling constant is (see appendix)
\begin{equation}
  \frac{\d\lambda}{\d\ln\nu(\Upsilon)}=\lambda-c\lambda^2
  \label{eq:dlambda_dl}
\end{equation}
with a non-negative coefficient
\begin{equation}
  c=\frac{\d\Pi_{pp}}{\d\nu(\Upsilon)}-\frac{\d\Pi_{ph}}{\d\nu(\Upsilon)}\geq0,
  \label{eq:dlambda_dl-coeff-c}
\end{equation}
where $\Pi_{pp}$ and $\Pi_{ph}$ are particle-particle and particle-hole polarization operators.

The scaling behavior of the system strongly resembles that of 1D
interacting electrons. Namely, exactly at the monkey saddle at $\mu=0$
the one-loop contribution to beta function vanishes, leaving a
critical theory with tree-level scaling only
\begin{equation}
\label{RG_lambda_T}
	\frac{\d\lambda}{\d\ln\nu(T)}=\lambda\quad(\mu=0,\forall T).
\end{equation}
This behavior is linked to an additional symmetry\cite{kapustin} that
arises exactly at the monkey saddle, and is a combination of
time-reversal transformation
$(\varepsilon,\vec{p})\rightarrow(-\varepsilon,-\vec{p})$ plus a
particle-hole transformation
$\psi^\dagger\rightleftharpoons\psi$. This symmetry is present only
for odd saddles with $\xi(-\vec{p})=-\xi(\vec{p})$ and is absent for
even saddles that have a dispersion that is invariant under spatial
inversion.

\begin{figure}[t]
	\begin{minipage}[h]{0.99\linewidth}
		\center{\includegraphics[width=1\linewidth]{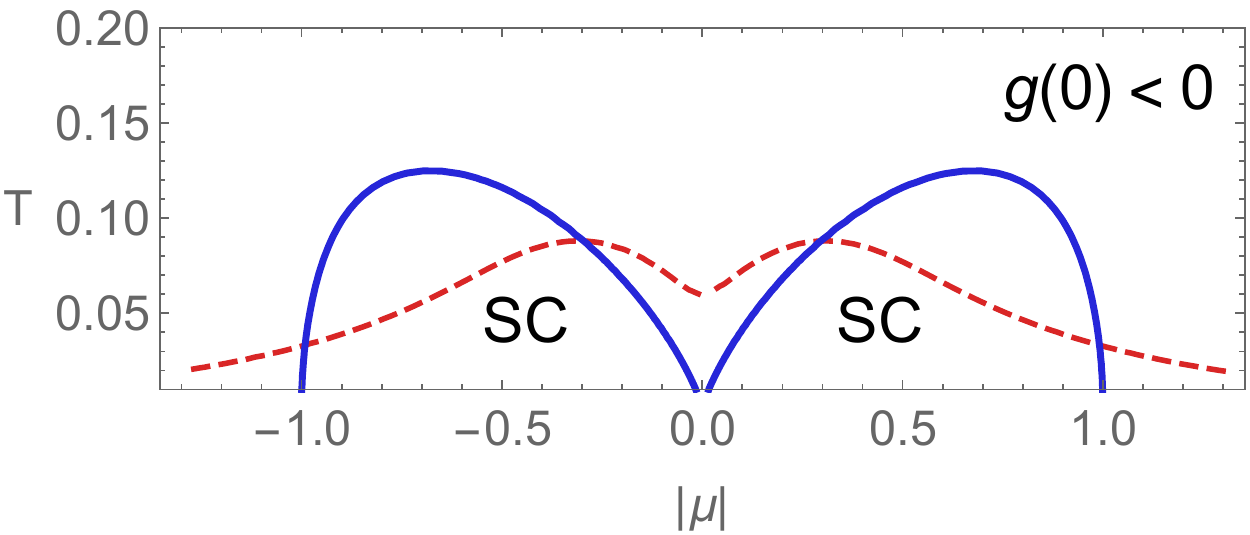}}
	\end{minipage}
	\caption{Phase diagram (blue solid line) for an isolated monkey saddle
		and attractive coupling constant. Critical chemical potential is
		determined by the equation $\abs{g_0}\nu(\mu_c)=2$ and the plot is
		given in units of $\mu_c$ for both temperature and chemical
		potential. Any odd saddle ($n=3,5,\dots$) has qualitatively same
		phase diagram, but the situation is different for even saddles
		($n=2,4,\dots$). Even case is illustrated with red dashed line for
		$n=4$.}
	\label{fig:RG_phase_diagram}
\end{figure}

At the same time, away from the monkey saddle 
\begin{equation}
\frac{\d\lambda}{\d\ln\nu(\mu)}=\lambda-\frac{1}{2}\lambda^2\quad(T\ll\abs{\mu}\neq0),
\end{equation}
and the system either flows to a non-trivial fixed point $\lambda=2$
for any positive initial coupling constant $\lambda_0>0$ or develops a
superconducting instability with $\lambda$ diverging as (for
$\lambda_0<0$)
\begin{equation}
\lambda(\mu)=\frac{\nu(\mu)g_0}{1+2g_0[\nu(\mu)-\nu_0]}
\simeq
\frac{3\mu_c}{2(\mu_c-\mu)}.
\end{equation}
Here $\nu_0$ and $g_0$ are the DoS and coupling constant at the
initial energy scale, while $\mu_c$ marks the energy scale
corresponding to the instability.
This leads to a non-BCS type of behavior for the critical energy scale
\begin{equation}
\mu_c,T_c\propto g_0^{\frac{n}{n-2}}\underset{(n=3)}{=}g_0^3.
\end{equation}
In fact, the one-loop RG equations can be integrated out for any
$\mu,T$ and the solution is equivalent to resummation of a leading
diagrammatic series in the language of Feynman diagrams. The resulting
expression for a dimensional coupling constant $g$ reads as
\begin{equation}
	g^{-1}|_{(\mu,T)}=\left(\Pi_{pp}-\Pi_{ph}\right)|_{(\mu,T)}+g_0^{-1},
\end{equation}
Thus, within a one-loop approximation, the phase transition line for
attractive interaction $g<0$ is determined by the equation
\begin{equation}
	g_0\left(\Pi_{pp}-\Pi_{ph}\right)|_{(\mu,T)}+1=0
\end{equation}
and the resulting phase diagram is given in
Fig.\ref{fig:RG_phase_diagram}.

As to the quasiparticle width, it is zero within the one-loop
approximation. A non-zero result can be obtained from a two-loop
diagram that yields a quasiparticle width at the monkey saddle
($\mu=0$) that signals non-Fermi-liquid behavior
\begin{equation}
	\Gamma\sim \lambda^2(T)\;T\propto T^{1/3},
\end{equation}
since for $\mu=0$ there is only a tree-level scaling and
$\lambda(T)=g\,\nu(T)\propto T^{-1/3}$ for an invariant value of the
dimensionful coupling constant $g$.  This implies that our analysis
breaks down at energy scales $T^*\sim \Gamma(T^*)$, or equivalently
when dimensionless coupling constant $\lambda(T^*)\gtrsim1$ becomes
too large.

The situation is the same for any odd saddle, $n=3,5,\dots$, but is
very different for even saddles. For even saddles there is no
cancellation of the one-loop contribution, so that $c\neq0$ at $\mu=0$
and the dimensionless coupling constant flows to a fixed point
$\lambda=1/c$ yielding marginal Fermi liquid behavior with decay rate
$\Gamma\sim T$. While this implies a dimensionless coupling constant
of order one, the existence of this fixed point could be justified
within $1/N$ expansion techniques.

\section{RG flow for bilayer graphene}
\label{sec:RG2}

In BLG there are two copies of the monkey saddle at the $K$ and
$K^\prime$ points, which are related by time-reversal symmetry, with
dispersions $\xi_\pm(\vec{p})=\pm\xi(\vec{p})$. The four-fermion
interaction now has three coupling constants:
\begin{align}
	\frac{g}{2}(\psi^\dagger\psi)^2=g_1(\psi^\dagger_{+ i}\psi^\dagger_{- j}\psi_{+ j}\psi_{- i}) + g_2(\psi^\dagger_{+i}\psi^\dagger_{- j}\psi_{- j}\psi_{+ i} ) 
	\nonumber
	\\
	+ \xcancel{g_3(\psi^\dagger_{+ i}\psi^\dagger_{+ j}\psi_{- j}\psi_{- i})} + g_4(\psi^\dagger_{\alpha \uparrow}\psi^\dagger_{\alpha \downarrow}\psi_{\alpha \downarrow}\psi_{\alpha \uparrow}),
\end{align}
where $i,j=\uparrow\downarrow$ indices stand for spin and $\alpha=\pm$
corresponds to $K/K'$ valley isospin, respectively. Our notation for
coupling constants is the same as in
Refs.~\onlinecite{furukawa,nandkishore}. The Umklapp $g_3$ coupling is
forbidden because the $K$ and $K^\prime$ points are inequivalent in
the sense of momentum conservation modulo reciprocal lattice vector,
$\vec{Q}=2\vec{p}_{KK^\prime}\not\simeq\vec{0}$.

There are now four polarization operators that drive the RG flow,
particle-particle and particle-hole at zero and $\vec{Q}$ momentum
transfer. We focus on BLG tuned exactly at the monkey saddle with both
critical voltage bias $\delta=1$ and chemical potential $\mu=0$.  The
relative roles of polarization operators are
\begin{align}
\label{d1}
	&d_0\equiv\frac{\d\Pi_{pp}(\vec{Q})}{\d\Pi_{pp}(\vec{Q})}=1,
	\, &d_2\equiv\frac{\d\Pi_{ph}(\vec{0})}{\d\Pi_{pp}(\vec{Q})}=1,
	\\
	\label{d2}
	&d_1\equiv\frac{\d\Pi_{ph}(\vec{Q})}{\d\Pi_{pp}(\vec{Q})}=3,
	\, &d_3\equiv\frac{\d\Pi_{pp}(\vec{0})}{\d\Pi_{pp}(\vec{Q})}=3.
\end{align}
Since $\Pi_{pp}(\vec{Q})\sim\nu(T)$, it is reasonable to define
dimensionless interaction constants as
$\lambda_i=g_i\Pi_{pp}(\vec{Q})$ and take $\d[\ln\Pi_{pp}(\vec{Q})]$
as RG time. This gives RG equations
\begin{align}
\label{bilayer_RG}
	\dot{\lambda}_1=&\lambda_1-6\lambda_1^2+2\lambda_1\lambda_4,
	\\
	\dot{\lambda}_2=&\lambda_2+2(\lambda_1-\lambda_2)\lambda_4-3\lambda_1^2,
	\\
	\dot{\lambda}_4=&\lambda_4+\lambda_1^2+2\lambda_1\lambda_2-2\lambda_2^2,
\end{align}
and the RG flow is in fact similar to that of the square
lattice\cite{furukawa} with parameters $d_i$ given by
Eqs.(\ref{d1},\ref{d2}) and one interaction channel turned off,
$g_3\equiv0$.

\begin{figure}[t]
	\begin{minipage}[h]{0.49\linewidth}
		\center{\includegraphics[width=1\linewidth]{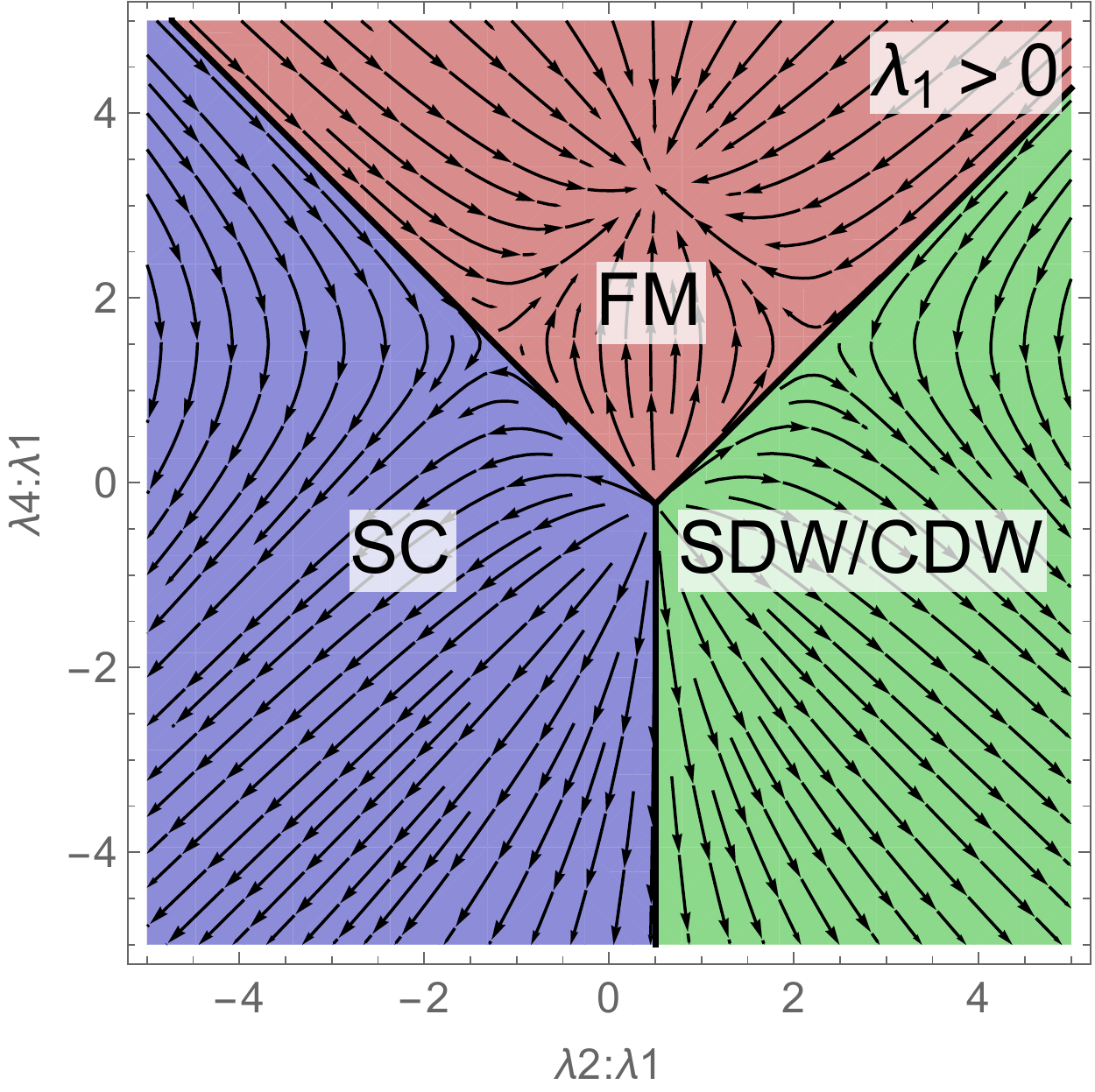}}
	\end{minipage}
	\hfill
	\begin{minipage}[h]{0.49\linewidth}
		\center{\includegraphics[width=1\linewidth]{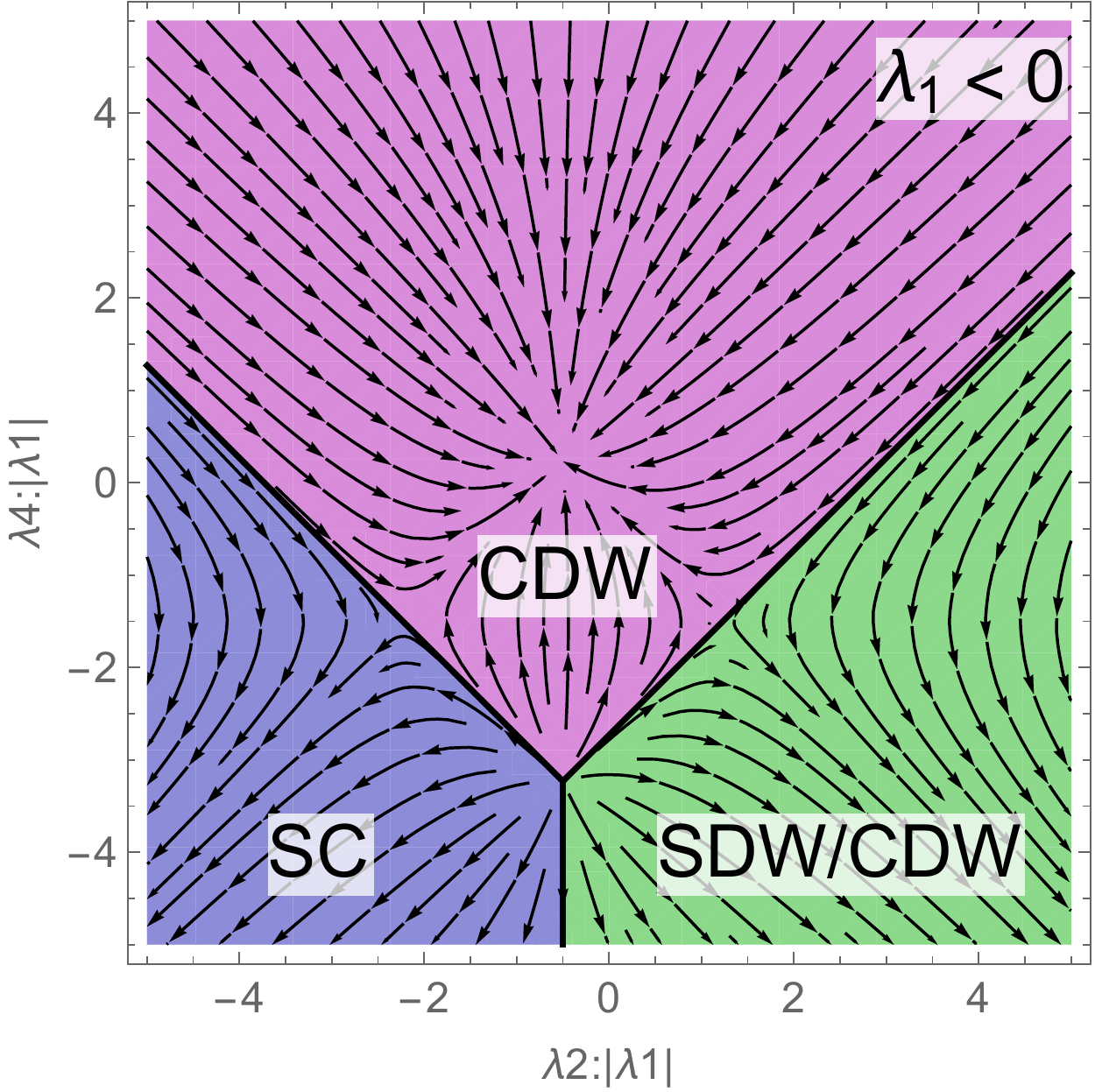}}
	\end{minipage}
	\caption{RG phase diagram showing a leading instability as a
		function of initial coupling constants. The figure on the left
		shows the case of positive $\lambda_1>0$, while the one on the
		right corresponds to $\lambda_1<0$. ($\lambda_1$ never changes
		sign under the RG flow.)  There are four possible instabilities:
		SC superconducting, ferromagnetic (FM), charge-density wave (CDW) and a competing spin/charge-density-wave
		(SDW/CDW). The Hubbard model initial conditions
		$\lambda_1=\lambda_2=\lambda_4>0$ lead to the development of FM
		instability.}
	\label{fig:BLG_phase_diagram}
\end{figure}

The crucial difference with the case of a single monkey saddle is that
the solution $\lambda_1=\lambda_2=0$ describing two decoupled saddles
is now always unstable.  The analysis of the RG flow is presented in
the appendix and it shows that there are four possible many-body
instabilities, $s$-wave superconducting (SC), ferromagnetic (FM), charge-density-wave (CDW) and a competing spin/charge-density-wave (SDW/CDW). However, only three instabilities,
SC, FM, and SDW/CDW are possible for initially repulsive interactions, as
is shown in a Fig. \ref{fig:BLG_phase_diagram}. For Hubbard model the
initial conditions correspond to all interaction constants being equal
and positive, $\lambda_i=(\lambda)_0>0$, and lead to FM phase.

\section{Conclusions}
\label{conclusions}

We studied the properties of electronic systems tuned to a monkey
saddle singularity, where the dispersion is $\propto
p_x^3-3p_xp_y^2$. We showed that such a situation occurs in a MLP
where three vH singularities merge. We showed that such a singular
point is accessible in BLG by controlling two parameters, the
interlayer bias voltage and the chemical potential. We identified a
number of experimentally accessible features associated with the
monkey saddle dispersion when the system is subject to a magnetic
field. The Landau level structure has a trademark behavior where
$E_m\propto (Bm)^{3/2}$, different from the behavior of both linearly
or quadratically dispersing systems. The oscillations of either
thermodynamic or transport properties with the applied magnetic field
(de Haas-van Alphen or Shubnikov-de Haas oscillations) contain a
signature tripling of the oscillation period when the Fermi energy
crosses the saddle point energy. This tripling, associated with the
topological transition between a single- and three-sheeted FS, can be
viewed as a smoking gun of the monkey saddle singularity.

Generically, the singular electronic dispersion in such MLP implies a
strong tendency towards development of many-body instabilities. We
found that the stronger divergence of the DoS in monkey saddle
singularities ($n=3$), as compared to the case of ordinary vH
singularities ($n=2$), brings about crucial simplifications in the
field theoretical analysis of the effect of interactions. We showed
that the theory for systems with higher order singularities ($n>2$) is
super-renormalizable. Thus, in contrast to the case of vH
singularities where an RG analysis requires two cut-off scales to
properly account for the singular and non-singular parts of the FS,
the analysis of higher order saddles requires no large momentum (UV)
cut-off, since there are only IR divergences, which are regularized by
temperature $T$ and chemical potential $\mu$.

Via an RG analysis of the super-renormalizable theory, we showed that
the non-interacting electron fixed point of a system with a single
monkey saddle is unstable to interactions, developing either a
superconducting instability or non-Fermi liquid behavior. We also
showed that the electronic lifetime depends crucially on the symmetry
of the dispersion, with odd and even saddles displaying
non-Fermi-liquid and marginal Fermi liquid behavior, respectively. For
BLG, which has two two non-nested monkey saddles at the $K$ and $K'$
points, we showed that interactions (depending on their nature) lead
to $s$-wave superconductivity, ferromagnetism, charge-density wave, or spin-density
wave.

The studies of MLP in electronic systems suggest an exciting link to
catastrophe and singularity theories. Namely, the monkey saddle could
be considered as a lattice-symmetry-restricted elliptical umbilic
elementary catastrophe $D_4^-$. Catastrophe theory may be a useful
language to classify the different possible singularities where FS
topology changes. The relevant classification at criticality is not
that of the FS topologies, but of the singularity itself. Controlling
the chemical potential and the interlayer bias voltage in BLG is a
clear example of how to engineer a catastrophe in an electronic
system, the monkey saddle. Crystalline symmetries may reduce the
possible types of catastrophes in the ADE classification that could be
realized in solid state systems. Which other singularities could occur
in electronic systems remains an open problem. However, our analysis
of the physical consequences of such singularities should be
applicable to other types of catastrophe in systems of electrons.

\section{Acknowledgements}

This work was supported in part by the Engineering
and Physical Sciences Research Council (EPSRC) Grant No. EP/M007065/1 (G.G.) and by the DOE Grant DEF-06ER46316 (C.C.).

\appendix

\section{RG analysis for an isolated monkey saddle}

\subsection{RG flow}

\begin{figure}[t]
	\center{\includegraphics[width=.8\linewidth]{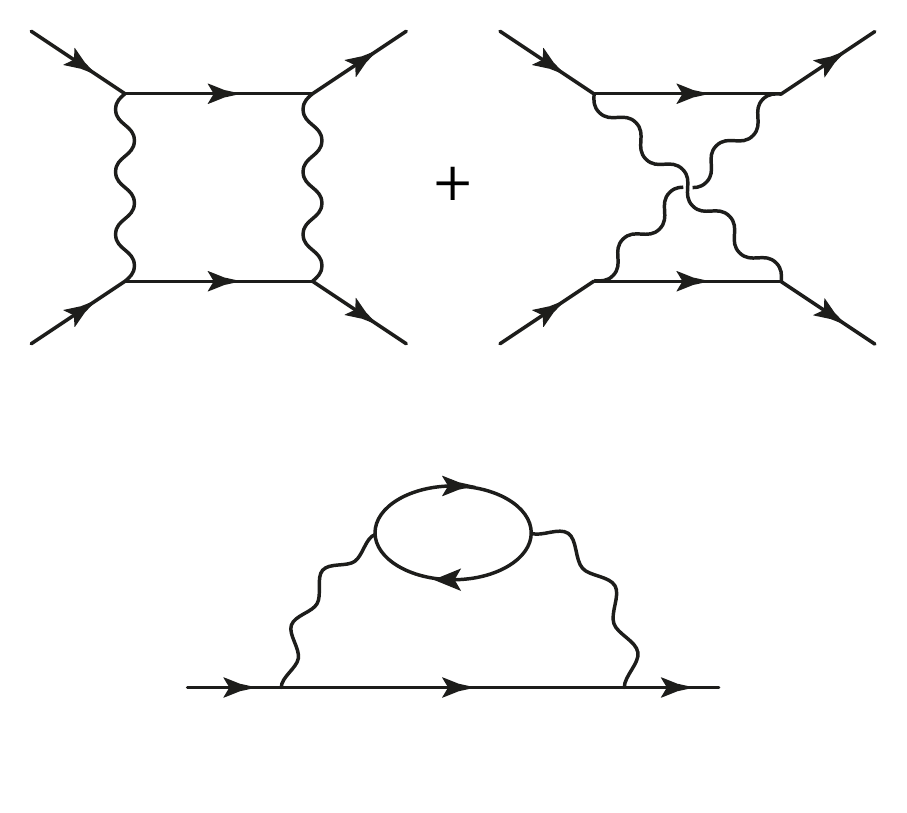}} 
	\caption{\textit{Top}: one-loop contribution to renormalization of the interaction constant for an isolated monkey saddle.
		\newline \textit{Bottom}: two-loop contribution to the quasiparticle decay rate.}
	\label{fig:appendix}
\end{figure}

The RG flow equation for the dimensionless interaction $\lambda$ constant is connected to renormalization of the dimensional coupling constant $g$ as
\begin{equation}
  \frac{\d\lambda}{\d\ln\nu}=\frac{\d(\nu g)}{\d\ln\nu}=\lambda+\nu^2\;\frac{\d g}{\d\nu}.
  \label{eq:A1}
\end{equation}
The one-loop renormalization of $g$ is given by two diagrams shown in
Fig.~\ref{fig:appendix} and yield
\begin{equation}
  \delta g= -g^2\Pi_{pp}(\mu,T)+g^2\Pi_{ph}(\mu,T).
  \label{eq:A2}
\end{equation}
Combining Eqs.~\eqref{eq:A1} and \eqref{eq:A2} we obtain the RG
equation for $\lambda$,
\begin{equation}
\frac{\d\lambda}{\d\ln\nu(\Upsilon)}=\lambda-c\lambda^2,\quad
c=\frac{\d\Pi_{pp}}{\d\nu(\Upsilon)}-\frac{\d\Pi_{ph}}{\d\nu(\Upsilon)}\geq 0
\;,
\end{equation}
presented in the main text as Eqs.~\eqref{eq:dlambda_dl} and~\eqref{eq:dlambda_dl-coeff-c}.

The polarization operators are defined as
\begin{align}
	\Pi_{ph}(\vec{q},\mu,T)=&-T\sumint_{l,\vec{p}}G(i\varepsilon_{l},\vec{p}+\vec{q})G(i\varepsilon_{l},\vec{p}),
\\
	\Pi_{pp}(\vec{q},\mu,T)=&T\sumint_{l,\vec{p}}G(i\varepsilon_{l},\vec{p}+\vec{q})G(-i\varepsilon_{l},-\vec{p}),
\end{align}
and the particle-hole polarization operator can be evaluated to be
\begin{align}
\Pi_{ph}=&-T\int_{\vec{p}}\sum_l\frac{1}{i\varepsilon_l-\xi_{\vec{p}+\vec{q}}+\mu}\frac{1}{i\varepsilon_l-\xi_{\vec{p}}+\mu}
\\
=&\;\frac{1}{2}\int_{\vec{p}}\frac{f(\xi_{\vec{p}+\vec{q}}-\mu)-f(\xi_{\vec{p}}-\mu)}{\xi_{\vec{p}+\vec{q}}-\xi_{\vec{p}}}
\\
\underset{\vec{q}\rightarrow0}{=}&\;\frac{1}{2}\int\nu(\xi)\,f^\prime(\xi-\mu)\;\d\xi
\;,
\end{align}
where $f(\xi)=\tanh\xi/2T$.

Similarly, the particle-particle polarization operator is
\begin{align}
\Pi_{pp}=&\frac{1}{2}\int_{\vec{p}}\frac{f(\xi_{\vec{p}}-\mu)+f(\xi_{-\vec{p}}-\mu)}{\xi_{\vec{p}}+\xi_{-\vec{p}}-2\mu}
\\
=&\frac{1}{2}\int\nu(\xi)\,\frac{f(\xi+\mu)-f(\xi-\mu)}{2\mu}\;\d\xi
\;.
\end{align}

The difference of polarization operators that drives RG flow has the
following asymptotic behavior:
\begin{equation}
\Pi_{pp}-\Pi_{ph}=
\begin{cases}
0 & \mu =0,\,T\neq0
\\
\dfrac{1}{2}\nu(\mu) & T = 0,\,\mu\neq0
\end{cases},
\end{equation}
where the cancellation at $\mu=0$ in fact holds for any external frequency and momentum.

The chemical potential also has a correction due to a Hartree-type
diagram,
\begin{equation}
\delta\mu=g\sumint_{l,\vec{p}} G(i\varepsilon_l,\vec{p}),
\end{equation}
corresponding to the shift in the monkey saddle's Fermi energy. (This
contribution is the equivalent of the fluctuational renormalization of
the critical temperature in thermodynamic phase transitions.)

Finally, we point out that the cancellation of the one-loop
contribution at $\mu=0$ is a feature specific to odd saddles. For an
$n$-th order saddle with a dispersion $\xi=p^n\cos n\phi$ the DoS
behaves as $\nu(\varepsilon)\propto\varepsilon^{-(n-2)/n}$, while the
polarization operators behave as
\begin{equation}
	\Pi_{pp}-\Pi_{ph}=
	\begin{cases}
	\frac{1+(-1)^n}{n}C_n\nu(\mu) & \mu =0,\,T\neq0
	\\
	\frac{n-2}{2}\nu(\mu) & T = 0,\,\mu\neq0
	\end{cases},
\end{equation}
with a (positive) numerical constant 
\begin{equation}
	\begin{split}
	C_n=&\int_0^\infty \d x\,x^{-(n-2)/n}(2\cosh^2(x/2))^{-1}
	\\
	=&2(2^{2/n}-1)\Gamma\left(2-\frac{2}{n}\right)\left[-\zeta\left(1-\frac{2}{n}\right)\right].
	\end{split}
\end{equation}
As we mentioned in the main text, this difference leads to
non-Fermi-liquid and marginal Fermi liquid behavior for odd and even
saddles, respectively.

\subsection{Quasiparticle decay rate}

The quasiparticle decay rate is related to the imaginary part of the
electron self-energy, which can be written (using the real-time
Keldysh technique) as
\begin{equation}
	\begin{split}
	\Delta\Sigma(\varepsilon,\vec{p})=&-\int_{\omega,\vec{q}}\left[B(\omega)+f(\varepsilon-\omega)\right]\times
	\\
	&\times\Delta G(\varepsilon-\omega,\vec{p}-\vec{q})\Delta L(\omega,\vec{q})
	\\
	=&-i\int_{\vec{q}}\left[B(\varepsilon-\xi_{\vec{p}-\vec{q}})+f(\xi_{\vec{p}-\vec{q}})\right]\times
	\\
	&\times
	\Delta L(\varepsilon-\xi_{\vec{p}-\vec{q}},\vec{q}),
	\end{split}
\end{equation}
where $B(x)=\coth (x/2T)$ and $f(x)=\tanh(x/2T)$ are bosonic and
fermionic distribution functions, $L$ is an interaction propagator,
and $\Delta(\dots) =(\dots)^R-(\dots)^A$ stands for the difference
between retarded and advanced components. The interaction propagator
within the one-loop approximation is essentially
\begin{equation}
\begin{split}
	\Delta\Sigma(\varepsilon,\vec{p})=&-ig^2\int_{\vec{k},\vec{q}}
	\delta(\varepsilon+\xi_{\vec{k}+\vec{q}-\vec{p}}-\xi_{\vec{q}}-\xi_{\vec{k}})\times
	\\
	&\times\Big(
		f(\xi_{\vec{q}})[f(\xi_{\vec{k}})-f(\xi_{\vec{k}}+\xi_{\vec{q}}-\varepsilon)]+
		\\
		&
		+1-f(\xi_{\vec{k}})f(\xi_{\vec{k}}+\xi_{\vec{q}}-\varepsilon)
	\Big),
\end{split}
\end{equation}
where we made use of the relation between equilibrium distribution
functions $[f(x+y)-f(x)]B(y)=1-f(x+y)f(x)$, and redefined integration
variables $\vec{k},\vec{q}$. This equation is essentially a statement
of Fermi's golden rule.

Rescaling momenta as $(\vec{k},\vec{q})\rightarrow
T^{1/3}(\vec{k},\vec{q})$ we see that the quasiparticle width at the
monkey saddle for zero chemical potential and zero external frequency
and momenta behaves as
\begin{equation}
	\Gamma=\frac{i}{2}\Delta\Sigma(0,\vec{0})\Big|_{\mu=0}\sim [\underbrace{\nu(T)g}_{\lambda(T)}]^2T\propto T^{1/3}.
\end{equation}

On the other hand, for non-zero chemical potential we find regular
Fermi-liquid-like behavior\cite{sachdev_book},
\begin{equation}
\Gamma\sim \lambda^2(\mu)\frac{\varepsilon^2}{\mu}\ln\frac{\mu}{\varepsilon},\quad T\ll\varepsilon\ll\abs{\mu}.
\end{equation}

\section{RG analysis for BLG}

\subsection{Polarization operators}

In BLG there are two additional polarization operators, with non-zero momentum transfer $\vec{Q}$
\begin{align}
\Pi_{ph}(\vec{Q},\mu,T)=&-T\sumint_{l,\vec{p}}G(i\varepsilon_{l},\vec{p})G(i\varepsilon_{l},\vec{Q}+\vec{p}),
\\
\Pi_{pp}(\vec{Q},\mu,T)=&T\sumint_{l,\vec{p}}G(i\varepsilon_{l},\vec{p})G(-i\varepsilon_{l},\vec{Q}-\vec{p}).
\end{align}
Once calculated, they yield
\begin{align}
	\Pi_{ph}(\vec{Q})=&
	\frac{1}{2}\int\nu(\xi)\,\frac{f(\xi-\mu)-f(-\xi-\mu)}{2\xi}\;\d\xi,
	\\
	\Pi_{pp}(\vec{Q})=&
	\frac{1}{2}\int\nu(\xi)\,\frac{f(\xi+\mu)}{\xi+\mu}\;\d\xi.
\end{align}

In this paper we focus on the case when the system is tuned to the
monkey saddle, $\mu=0$, where
\begin{align}
\label{appendix:pi1}
	\Pi_{ph}(\vec{0})&=\Pi_{pp}(\vec{0})=C_3\;\nu(T),\quad&(\mu=0)
	\\
	\label{appendix:pi2}
	\Pi_{ph}(\vec{Q})&=\Pi_{pp}(\vec{Q})=3C_3\;\nu(T),\quad&
\end{align}
with numerical constant 
\begin{equation}
	C_3=\int_0^\infty \d x\,x^{-1/3}\frac{1}{2\cosh^2(x/2)}=1.14.
\end{equation}

\subsection{RG equations}

The RG flow equations for a square lattice with two hot spots were
derived in Ref. \onlinecite{furukawa}. These equations are very
general and in their infinitesimal form, after an elementary RG step,
they give
\begin{equation}
	\begin{split}
	\delta{g}_1=&2g_1(g_2-g_1)\delta\Pi_{ph}(\vec{Q})+2g_1g_4\delta\Pi_{ph}(\vec{0})
	\\
	&-2g_1g_2\delta\Pi_{pp}(\vec{Q}),
	\\
	\delta{g}_2=&(g_2^2+g_3^2)\delta\Pi_{ph}(\vec{Q})+2(g_1-g_2)g_4\delta\Pi_{ph}(\vec{0})
	\\
	&-(g_1^2+g_2^2)\delta\Pi_{pp}(\vec{Q}),
	\\
	\delta{g}_3=&-2g_3g_4\delta\Pi_{pp}(\vec{0})+2(2g_2-g_1)g_3\delta\Pi_{ph}(\vec{Q})
	\\
	\delta{g}_4=&-(g_3^2+g_4^2)\delta\Pi_{pp}(\vec{0})
	\\
	&+(g_1^2+2g_1g_2-2g_2^2+g_4^2)\delta\Pi_{ph}(\vec{0}).
	\end{split}
\end{equation}
In the case of BLG there is no Umklapp scattering between $K$ and
$K^\prime$ points, and thus we set $g_3\equiv0$. The coupling
constants $g_i$ are dimensionful, but we introduce dimensionless
coupling constants as follows. Since $\Pi_{pp}(\vec{0})\propto\nu$
(see Eq.~\ref{appendix:pi1}), it is appropriate and convenient to
define the dimensionless constants as
$\lambda_i=\Pi_{pp}(\vec{0})g_i$, and take $\d\ln\Pi_{pp}(\vec{0})$
for RG time $\d s$:
\begin{equation}
	\begin{split}
	\dot{\lambda}_1=&\lambda_1+2d_1\lambda_1(\lambda_2-\lambda_1)+2d_2\lambda_1\lambda_4-2d_3\lambda_1\lambda_2,
	\\
	\dot{\lambda}_2=&\lambda_2+d_1\lambda_2^2+2d_2(\lambda_1-\lambda_2)\lambda_4-d_3(\lambda_1^2+\lambda_2^2),
	\\
	\dot{\lambda}_4=&\lambda_4-d_0\lambda_4^2+d_2(\lambda_1^2+2\lambda_1\lambda_2-2\lambda_2^2+\lambda_4^2),
	\end{split}
\end{equation}
where for the sake of generality we introduced an additional parameter
$d_0$. Parameters $d_i$ are defined in the main text by
Eqs. \ref{d1},\ref{d2} and their explicit numerical values follow from
Eqs. \ref{appendix:pi1},\ref{appendix:pi2}. This scheme gives the RG
flow presented in the main text,
\begin{align}
	\label{appendix:lambda1}
	\dot{\lambda}_1=&\lambda_1-6\lambda_1^2+2\lambda_1\lambda_4,
	\\
	\dot{\lambda}_2=&\lambda_2+2(\lambda_1-\lambda_2)\lambda_4-3\lambda_1^2,
	\\
	\dot{\lambda}_4=&\lambda_4+\lambda_1^2+2\lambda_1\lambda_2-2\lambda_2^2.
\end{align}

At the brink of a many-body instability the coupling constants diverge
as
\begin{equation}
	\lambda_i=\frac{\lambda_i^{(0)}}{s_c-s},
\end{equation}
where $s_c$ is a critical RG time corresponding to the instability. By
seeking solutions of this form we get a system of algebraic equations
\begin{equation}
	\begin{split}
	\lambda_1^{(0)}=&-6\left(\lambda_1^{(0)}\right)^2+2\lambda_1^{(0)}\lambda_4^{(0)},
	\\
	\lambda_2^{(0)}=&2\left(\lambda_1^{(0)}-\lambda_2^{(0)}\right)\lambda_4^{(0)}-3\left(\lambda_1^{(0)}\right)^2,
	\\
	\lambda_4^{(0)}=&\left(\lambda_1^{(0)}\right)^2+2\lambda_1^{(0)}\lambda_2^{(0)}-2\left(\lambda_2^{(0)}\right)^2.
	\end{split}
\end{equation}
This system has the following four stable solutions
\begin{align}
	\lambda_1:\lambda_2:\lambda_4=&2:1:(3+\sqrt{12}) &\text{(FM)}
	\\
	=&0:1:(-1) &\text{([S/C]DW)}
	\\
	=&(-2):(-1):(\sqrt{12}-3)&\text{(CDW)}
	\\
	=&0:(-1):(-1) &\text{(SC)}
\end{align}
that correspond to ferromagnetic (FM), competing spin- and charge-density-wave ([S/C]DW),
charge-density-wave (CDW) and $s$-wave SC instabilities respectively.

The nature of instabilities is identified with the help of the
susceptibilities calculated in
Refs.\cite{furukawa,dzyaloshinskii,lederer}. Susceptibilities to different order
parameters diverge as $\chi_j\propto(s_c-s)^{\alpha_j}$, so the
leading instability is the one with the most negative value of
$\alpha_j$, given by
\begin{align}
\alpha_{sP_\vec{Q}}=&2\lambda_4^0
\\
\alpha_{s_\pm P_\vec{Q}}=&2\lambda_4^0
\\
\alpha_{CDW}=&6(2\lambda_1^0-\lambda_2^0)
\\
\alpha_{SDW}=&-6\lambda_2^0
\\
\alpha_{\text{spin}}=&-2(\lambda_1^0+\lambda_4^0)
\\
\alpha_{\text{charge}}=&2(-\lambda_1^0+2\lambda_2^0+\lambda_4^0)
\\
\alpha_{sP}=&6(-\lambda_1^0+\lambda_2^0)
\end{align}
for finite momentum $s$-wave and $s_\pm$-wave superconducting, charge density wave, spin
density wave, ferromagnetic (uniform spin), uniform charge ($\kappa$), and $s$-wave superconducting instabilities respectively.

Susceptibilities can be calculated by studying renormalization of test vertices\cite{nandkishore}. The first group of instabilities correspond to uniform densities with a test Lagrangian density
\begin{align}
\delta\mathcal{L}=\sum_{i=\uparrow\downarrow}\sum_{\alpha=+-}n_{i\alpha}\psi^\dagger_{i\alpha}\psi_{i\alpha},
\end{align}
where renormalization of test vertices $n_{i\alpha}$ within one-loop approximation is given by equation
\begin{align}
\label{appendix:test}
\frac{\d}{\d s}
\begin{pmatrix}
n_{+\uparrow}
\\
n_{+\downarrow}
\\
n_{-\uparrow}
\\
n_{-\downarrow}
\end{pmatrix}
=
d_2
\begin{pmatrix}
0 & -\lambda_4 &\lambda_1-\lambda_2 & - \lambda_2
\\
-\lambda_4 & 0 & -\lambda_2 & \lambda_1-\lambda_2
\\
\lambda_1-\lambda_2 & -\lambda_2 & 0 & -\lambda_4
\\
-\lambda_2 & \lambda_1-\lambda_2 & -\lambda_4 & 0
\end{pmatrix}
\begin{pmatrix}
n_{+\uparrow}
\\
n_{+\downarrow}
\\
n_{-\uparrow}
\\
n_{-\downarrow}
\end{pmatrix}
\end{align}
and susceptibilities are equal to $\alpha=-2\gamma$, where $\gamma$ is an eigenvalue of (\ref{appendix:test}). Solving for eigensystem of (\ref{appendix:test}) we find four instabilities with susceptibilities
\begin{align}
\alpha_{\text{spin}}=&-2(\lambda_1^0+\lambda_4^0),
\\
\alpha_{\text{charge}}=&2(-\lambda_1^0+2\lambda_2^0+\lambda_4^0),
\\
\alpha_\text{{valley}}=&2(\lambda_1^0-2\lambda_2^0+\lambda_4^0),
\\
\alpha_{\text{spin-valley}}=&2(\lambda_1^0-\lambda_4^0).
\end{align}
The second group of instabilities is  a charge- and spin-density waves,
	\begin{align}
	\delta\mathcal{L}=\sum_{i=\uparrow\downarrow}n_{\vec{Q}i}\psi^\dagger_{-i}\psi_{+i}+\text{h.c.}
	\end{align}
	\begin{align}
	\frac{\d}{\d s}
	\begin{pmatrix}
	n_{\vec{Q}\uparrow}
	\\
	n_{\vec{Q}\downarrow}
	\end{pmatrix}
	=
	d_1
	\begin{pmatrix}
	\lambda_2-\lambda_1 & -\lambda_1
	\\
	-\lambda_1 & \lambda_2-\lambda_1
	\end{pmatrix}
	\begin{pmatrix}
	n_{\vec{Q}\uparrow}
	\\
	n_{\vec{Q}\downarrow}
	\end{pmatrix},
	\end{align}
	\begin{align}
	\alpha_{CDW}=&6(2\lambda_1^0-\lambda_2^0),
	\\
	\alpha_{SDW}=&-6\lambda_2^0.
	\end{align}
The third group represents superconducting $s$- and $s_\pm$-wave instabilities,
	\begin{align}
	\delta\mathcal{L}=\Delta_{1}\psi^\dagger_{+\uparrow}\psi^\dagger_{-\downarrow}+\Delta_{2}\psi^\dagger_{-\uparrow}\psi^\dagger_{+\downarrow}+\text{h.c.},
	\end{align}
	\begin{align}
	\frac{\d}{\d s}
	\begin{pmatrix}
	\Delta_{1}
	\\
	\Delta_{2}
	\end{pmatrix}
	=
	d_3
	\begin{pmatrix}
	-\lambda_2 & -\lambda_1
	\\
	-\lambda_1 & -\lambda_2
	\end{pmatrix}
	\begin{pmatrix}
	\Delta_{1}
	\\
	\Delta_{2}
	\end{pmatrix},
	\end{align}
	\begin{align}
	\alpha_{sP}=&6(\lambda_2^0+\lambda_1^0),
	\\
	\alpha_{s_{\pm}P}=&6(\lambda_2^0-\lambda_1^0).
	\end{align}
Finally, the last group corresponds to finite momentum superconductivities,
	\begin{align}
	\delta\mathcal{L}=\Delta_{s\vec{Q}+}\psi^\dagger_{+\uparrow}\psi^\dagger_{+\downarrow}+\Delta_{s\vec{Q}-}\psi^\dagger_{-\uparrow}\psi^\dagger_{-\downarrow}+\text{h.c.},
	\end{align}
	\begin{align}
	\frac{\d}{\d s}
	\begin{pmatrix}
	\Delta_{s1}
	\\
	\Delta_{s2}
	\end{pmatrix}
	=
	d_0
	\begin{pmatrix}
	-\lambda_4 & 0
	\\
	0 & -\lambda_4
	\end{pmatrix}
	\begin{pmatrix}
	\Delta_{s1}
	\\
	\Delta_{s2}
	\end{pmatrix},
	\end{align}
	\begin{align}
	\alpha_{sP_\vec{Q}}=&\lambda_4,
	\\
	\alpha_{s_{\pm}P_\vec{Q}}=&\lambda_4.
	\end{align}

Going back to the analysis of RG flow (\ref{appendix:lambda1}), since $\lambda_1$ cannot change sign (RHS for $\dot{\lambda}_1$ is equal to zero when $\lambda_1=0$), it
is convenient to analyze the RG flow in $y_2=\lambda_2/\lambda_1$ {\it
  vs.}  $y_4=\lambda_4/\lambda_1$ coordinates,
\begin{align}
	\dot{y}_2=&\lambda_1\left(-3+6y_2+2y_4-4y_2y_4\right),
	\\
	\dot{y}_4=&\lambda_1\left(1+2y_2+6y_4-2(y_2^2+y_4^2)\right).
\end{align}
We can then reparametrize the RG flow eliminating $\lambda_1$ to get a
system of equations
\begin{align}
y_2^\prime=&-3+6y_2+2y_4-4y_2y_4,
\\
y_4^\prime=&1+2y_2+6y_4-2(y_2^2+y_4^2),
\end{align}
that can be solved exactly in the coordinates $y_\pm$,
\begin{equation}
y_\pm=(y_4-3/2)\pm (y_2-1/2):\quad y_\pm^\prime=6-y_\pm^2.
\end{equation}
This allows us to identify all phases and phase boundaries on the
$y_2y_4$ plane. Thus, the plot in $\lambda_2/\lambda_1$ {\it vs.}
$\lambda_4/\lambda_1$ coordinates explicitly shows the fate of the
system for different initial coupling
constants. Fig.~\ref{fig:BLG_phase_diagram}~(left) shows the phase
diagram of RG flow for $\lambda_1>0$. FM, SC or competing [S/C]DW instabilities are
possible with phase boundaries
\begin{align}
	\lambda_2-\lambda_1/2=0, &\quad(\text{SC/SDW})
	\\
	\lambda_2+\lambda_4-(2-\sqrt{3})\lambda_1=0, &\quad(\text{FM/SC})
	\\
	\lambda_2-\lambda_4-(\sqrt{3}-1)\lambda_1=0, &\quad(\text{FM/[S/C]DW})
\end{align}
and the lines cross at the point
\begin{equation}
	\lambda_1:\lambda_2:\lambda_4=2:1:(3-\sqrt{12}).
\end{equation}

For negative values $\lambda_1<0$ we get options of SC, CDW, and [S/C]DW and Fig.~\ref{fig:BLG_phase_diagram}~(right). The phase boundaries are now
\begin{align}
\lambda_2+\abs{\lambda_1}/2=0, &\quad(\text{SC/[S/C]DW})
\\
\lambda_2+\lambda_4-(2-\sqrt{3})\abs{\lambda_1}=0, &\quad(\text{SC/CDW})
\\
\lambda_2-\lambda_4-(\sqrt{3}-1)\abs{\lambda_1}=0, &\quad(\text{CDW/[S/C]DW})
\end{align}
crossing at the point
\begin{equation}
	\abs{\lambda_1}:\lambda_2:\lambda_4=2:1:(3-\sqrt{12}).
\end{equation}

\bibliography{monkey}

\end{document}